\documentclass[12pt,a4paper]{article}
\usepackage{amsmath}
\usepackage{graphicx}
\usepackage[colorlinks=true,linkcolor=blue,urlcolor=blue,citecolor=blue]{hyperref}
\usepackage{cite}

\usepackage{IEEEtrantools}

\title{Properties of Tensor Hermite Polynomials}
\author{Parul Maheshwari$^{1,2}$, Gautam Mukhopadhyay$^1$,
\\ Siddhartha SenGupta$^2$
\\ \small $^1$Indian Institute of Technology Bombay, Mumbai.
\\ \small $^2$TCS Research Lab, Mumbai.
\\ \small parulm@iitb.ac.in
}
\date{}

\begin{document}

\maketitle

\begin{center}
\section*{Abstract}
\end{center}

A description of Orthogonal Tensor Hermite Polynomials in 3-D is presented. These polynomials, as introduced by Grad in 1949 \cite{1}, can be used to obtain a series solution to the Boltzmann Transport Equation. The properties that are explored are scaling, translation and rotation. Order 6 Hermite Tensors are studied while obtaining the rotation relations. From the scaling of the independent variables of particle velocities, a criterion on temperature is obtained which implies that the equation can be applied to binary gas mixtures only if the temperature of the hotter constituent is less than four times that of the cooler one. This criterion and other properties of the tensor hermite polynomials obtained in this paper can be used to study gas dynamics in the thermosphere.


\vspace{2pc}

\section{Introduction}

In non equilibrium statistical mechanics, the Boltzmann Transport Equation (BTE) is a widely used tool for describing the transport properties of a classical thermodynamic system. Since this equation is applicable mainly to low density systems for which binary collisions dominate, it gives, for example, the most suitable way of analysing the composition and dynamics of the upper atmosphere of the earth. In this paper, the properties of the solution of this equation are explored in a manner to be of value primarily in the study of thermosphere. The Boltzmann equation, when applied to a gas mixture like the one in consideration in upper atmosphere, leads to many difficult situations. The general interactions are varied and complex. Moreover, highly non-equilibrium conditions are found, for example, at high latitudes in the thermosphere, where the ions move at extremely high speeds compared to the neutral atoms due to the presence of magnetospheric electric fields. Another example can be the case of exobase (lower boundary of the exosphere) where the extremely sparse population fails to give a sufficiently high rate of collisions to justify the applicability of the collision term of the BTE. In this case, one would need a way to find whether or not the equation is valid. In order to address situations like these, the ordinary collision integrals used to simplify the BTE, fail to provide an advantageous insight in the situation due to the insurmountable complexity of the integrals.
The methods previously equipped to attempt at solving the BTE include the eigenvalue approach of Chapman and Cowling \cite{2}, simplification of the collision term by Chapman and Enskog \cite{3} and orthogonal polynomial method by Grad \cite{4}. Of these, the method used by Grad which primarily exploits the Tensor Hermite Polynomials to obtain a series solution to the transport equation, presents explorable prospects. 
\\In this paper, we study the Tensor Hermite Polynomials and their properties to understand their applicability in areas other than dynamics of upper atmosphere involving applications in computational fluid dynamics, study of magneto plasma in tokomak for fusion research, dynamics in the core of a nuclear reactor, etc.
The basic work on Tensor Hermite Polynomials was done by Grad in 1949 \cite{1} introducing the need for them. He introduced them as the basis for complete set of orthonormal polynomials in N variables in tensorial form. As mentioned in his work, this set of scalar orthonormal polynomials often lacks symmetry and an advantage is gained by expressing these polynomials in tensor invariant notation. Other than Grad, work on these tensor polynomials has been done by Viehland \cite{5} to solve the BTE for trace amounts of ions in dilute gases and by Knio \cite{6} to study the uncertainty propagation in computational fluid dynamics. 
\\In this paper, three major aspects of the tensor hermite polynomials are studied. In the first one, the effects of scaling are examined. This directly provides a mathematical criterion governing the ambient conditions that must be satisfied for justifying the applicability of the collision integral term of the BTE in gas dynamical procedures. The second studies the change in Hermite polynomials under a translation of axes. The third one inspects the polynomials under rotation of axes from particle velocity coordinates to the centre-of-mass and relative velocity coordinates. This results in tensors of order 6 which are required to evaluate the collision integrals for a two component system of low density gas mixture. 


\section{Tensor polynomials: series solution to Boltzmann Equation}

The tensorial hermite polynomials can be written iteratively as (\ref{app:A3}): 
\begin{equation}  \label{eq:eq1}
\boldsymbol{H}_{n+1}(z)=\frac{1}{(n+1)!}S_{(n+1,3)}[\boldsymbol{H}_n(z)\boldsymbol{H}_1(z)-2n\boldsymbol{H}_{n-1}(z)\underline{\underline{I}}],
\end{equation}
where the Permutation operator $S_{(n,3)}$ is explained in \ref{app:A2}. Hence, the first few polynomials are
\begin{flalign}
& \boldsymbol{H}_0 = 1, \label{eq:2}
\\& \boldsymbol{H}_1 = 2\underline{z}, \label{eq:3}
\\& \boldsymbol{H}_2 = 4\underline{z}_i\underline{z}_j - 2\underline{\underline{\delta}}_{ij}, \label{eq:4}
\\& \boldsymbol{H}_3 = 8\underline{z}_i\underline{z}_j\underline{z}_k - 4(\underline{z}_i\underline{\underline{\delta}}_{jk} + \underline{z}_j\underline{\underline{\delta}}_{ki} + \underline{z}_k\underline{\underline{\delta}}_{ij}). \label{eq:5}
\end{flalign}
The entities stated in bold are tensors. The rank of these entities are denoted by the number of underlines below it or by the numerical subscript. These polynomials differ from the tensor hermite polynomials introduced by Grad \cite{1} described in \ref{app:A1} and are related to the above definition.
\\These polynomials are orthonormal and symmetric. Hence, any scalar function, for example, the distribution function of the Boltzmann transport equation $f(\underline{z})$, can be expanded in terms of these Hermite Polynomials
\begin{equation} \label{eq:6}
f(\underline{z}) = \lim_{N \to \infty} \sum_{n=0}^N (\boldsymbol{a}_n,\boldsymbol{H}_n(\underline{z})),
\end{equation}
provided the integral
\begin{equation}  \label{eq:7}
\int_{-\infty}^{+\infty} d\underline{z}  \hspace{1 mm} w  \hspace{1 mm} {|f(\underline{z})|}^2 < \infty.
\end{equation}
Here, $\underline{z}$ is the dimensionless velocity vector such that:
\begin{equation} \label{eq:eq8}
\underline{z} = \underline{v}\sqrt{\frac{m}{2k_BT}},
\end{equation}
where $\underline{v}$ is the particle velocity, $m$ is its mass, $k_B$ is the Boltzmann constant and $T$ is the thermodynamic temperature. In equation \ref{eq:7}, $w$ is the weight, which, in this context is the normalised Maxwellian velocity distribution function displaced by an amount $\underline{v}_{av}$.
\begin{equation}
w(\underline{v}) = n {\left(\frac{m}{2\pi k_BT}\right)}^{3/2} e^{{-m {(\underline{v}-\underline{v}_{av})}^2}/{(2k_BT)}}.
\end{equation}

\noindent The $\boldsymbol{a}_n$'s in equation \ref{eq:6} are called the expansion coefficients. They can be obtained by exploiting the orthonormality of the Hermite polynomials described by 
\begin{equation}
\int_{-\infty}^\infty{  w  \hspace{1 mm} \boldsymbol{H}_m(\boldsymbol{z}) (\boldsymbol{a}_n, \boldsymbol{H}_n(\boldsymbol{z})) d\underline{\boldsymbol{z}}} = \boldsymbol{a}_m \delta_{mn}.
\end{equation} \label{eq:9}
\\The notation $(\textbf{A},\textbf{B})$ represents the scalar product of the tensors $\textbf{A}$ and $\textbf{B}$ (\ref{app:A4}).


\section{Transformation of variables}

\subsection{Scaling}
In this section we explore the possibility of conversion from one type of functional dependence to another where the other one is scaled with respect to $\underline{z}$.
Consider the relation,
\begin{equation} \label{eq:eq10}
\underline{z}' = \alpha (\underline{z} - \underline{z}_0),
\end{equation}
where $\underline{z}_0$ and $\alpha$ are constants. Henceforth, summation over repeated indices is implied unless stated otherwise.
Considering the general functional dependence of the form, $f(z)$ can be written in terms of $\underline{z}$ as \cite{8},
\begin{equation}
f(z) = f_{0} \enspace e^{-(\underline{z},\underline{z})} \sum_{n=0}^N(\boldsymbol{a}_n,\boldsymbol{H}_n(\underline{z})).
\end{equation} \label{eq:11}
\\If the same function has to be equivalently representable in terms of $\underline{z}'$, assuming the same functional dependence, through the expression:
\begin{equation}
f(z) = f_{0}' \enspace e^{-(\underline{z}',\underline{z}')} \sum_{n=0}^N(\boldsymbol{a}_n',\boldsymbol{H}_n(\underline{z}')),
\end{equation} \label{eq:12}
\\then the series $(\boldsymbol{a}_n',\boldsymbol{H}_n(\underline{z}'))$ should be convergent. Here, $f_0$ and $f'_0$ are constants. From the above expressions, the two series are related as:
\begin{equation}
(\boldsymbol{a}'_n,\boldsymbol{H}'_n)  = (f_{0}/f'_0) \enspace e^{[-(\underline{z},\underline{z}) + (\underline{z}',\underline{z}')]} (\boldsymbol{a}_n, \boldsymbol{H}_n),
\end{equation} \label{eq:13}
\\And the series in $\underline{z}'$ will converge if the following integral
\begin{flalign*}
\int_{-\infty}^\infty{||(\boldsymbol{a}'_n,\boldsymbol{H}'_n)||^2 \enspace e^{(\underline{z}',\underline{z}')} d\underline{z}'} 
 = (f_{0}/f'_0)^2 \alpha \int_{-\infty}^\infty{||(\boldsymbol{a}_n, \boldsymbol{H}_n)||^2 \enspace e^{[ (\underline{z}',\underline{z}') - 2(\underline{z},\underline{z})]} dz},
\end{flalign*}
exists. From the orthogonality of the hermite polynomials, we can write 
\begin{equation}  \label{eq:14}
||(\boldsymbol{a}_n, \boldsymbol{H}_n)|| = (\boldsymbol{A}_n,\boldsymbol{H}_n),
\end{equation}
where $\textbf{A}_n$ are functions of quadratic combinations of $\textbf{a}_n$'s. For a finite value of $n$, the expression in equation \ref{eq:14} will be finite since $\boldsymbol{H}_n(z)$ depends on finite powers of $z$.
Hence, the existence of the integral depends only on that of the integral of the exponent, i.e., 
\begin{equation}
\int_{-\infty}^\infty{e^{[ (\underline{z}',\underline{z}') - 2(\underline{z},\underline{z})]}\enspace dz} < \infty.
\end{equation} \label{eq:15}
\\This happens when $z^2$ has a negative coefficient in the exponent. By substituting for $z'$ from equation (\ref{eq:eq10}), we get the condition as $\alpha^2 < 2$.
\\From the expression of $z$ in equation (\ref{eq:eq8}), we can see that $\alpha^2 = T/T_S$ where $T_S$ is the temperature of the species with the scaled velocities and $T$ is the temperature of particles with unscaled velocity.
\\Hence the interchange of variables can be done if $2T_S>T$.
\\In studies of ion-neutral mixture in upper atmosphere, where $T_i \geq T_n$ ($T_i$ is the ion temperature and $T_n$ is the neutral atom temperature), interconversions to and from a temperature $T$ is possible when it falls in the range:
\begin{equation}
(1/2)T_i < T < 2 T_n.
\end{equation} \label{eq:16}
\\This also implies that for the Boltzmann Equation to hold, with the collision term, the ion temperature must be less than four times the neutral atom temperature.

\subsection{Translation}
Translation can be made from one frame of reference to another. There are generally three frames of reference: the inertial, the one moving with the entire gas, and the one moving with a particular component species. Let two translations be defined as:
\begin{flalign}
&\underline{z}_0 = \underline{z} - \underline{z}_{00}, \label{eq:17}
\\&\underline{z}_r = \underline{z} - \underline{z}_a, \label{eq:18}
\end{flalign}
where $\underline{z}_{00}$ is an arbitrary constant and $\underline{z}_a$ is the average value of $\underline{z}$.
\\The hermite polynomials under both the translations are inter-related through:
\begin{flalign}
&\boldsymbol{H}_n^0 = \frac{1}{n!}S_{(n,3)}[\sum_{p=0}^n (_p^n) \boldsymbol{H}_p^r [2(\underline{z}_a-\underline{z}_{00})]^{t(n-p)}], \label{eq:eq19}
\\&\boldsymbol{H}_p^r = \frac{1}{p!} S_{(p,3)}[\sum_{n=0}^p (_n^p) \boldsymbol{H}_n^0 [2(\underline{z}_{00} - \underline{z}_a)]^{t(p-n)}]. \label{eq:eq20}
\end{flalign}
In the above expressions, $\boldsymbol{H}_n^0$ and $\boldsymbol{H}_p^r$ represent the Hermite Polynomials in terms of $\underline{z}_0$ and $\underline{z}_r$ defined as per equation \ref{eq:17} and \ref{eq:18} respectively. Here, $(_p^n)$ is the usual binomial coefficient. These expressions can be obtained by the principle of mathematical induction and using equation (1) as explained in \ref{app:A7}. It should be realised that translation destroys the property of orthogonality.

\subsection{Rotation: Tensors of order 6}
In order to evaluate the collision term of Boltzmann Equation for the gas flow problem, one has to consider the collision integral and the collision cross section depends on the relative velocity of the collision partners \cite{9}. More often than not, this relative velocity can have a direction different from that of the species velocity. In that case, a rotation of axes makes the calculation a lot less cumbersome. 
\\Considering for example, $\underline{z}_{rs}$ and $\underline{z}_{rs'}$ as the space vectors and $f_s$ and $f_{s'}$ as the corresponding distribution functions for two species, the axes can be rotated to:
\begin{flalign}
&c_r = (m_s v_{rs} + m_{s'} v_{rs'})/ \sqrt{2k_BT(m_s+m_{s'})}, \label{eq:21}
\\&g_r = (v_{rs'} - v_{rs})\sqrt{\mu / 2k_BT}, \label{eq:22}
\\&v_{rs} = \sqrt{2k_BT/m_{s}} \enspace z_{rs}, \label{eq:23}
\\&v_{rs'} = \sqrt{2k_BT/m_{s'}} \enspace z_{rs'}, \label{eq:24}
\\& \mu = m_sm_{s'}/(m_s + m_{s'}), \label{eq:25}
\end{flalign}
where $c_r$ and $g_r$ are the rotated coordinates, $m_s$ and $m_{s'}$ are the molecular masses of the two species, $ k_B $ is the Boltzmann constant, $T$ is the temperature, $v_{rs}$ and $v_{rs'}$ are the species' velocities and $\mu$ is the reduced mass.
\\The rotation matrix can be expressed as:
\begin{flalign*}
&R_{1;s'sgc} = R_{1;gcs's} = \begin{bmatrix}
y \underline{\underline{I}} & y' \underline{\underline{I}} \\
y' \underline{\underline{I}} & -y \underline{\underline{I}}
\end{bmatrix},
\\&y^2 = \mu / m_{s}, \enspace y'^2 = \mu / m_{s'}.
\end{flalign*}
And the corresponding relations can be expressed as:
\begin{equation}
\begin{split}
&\begin{bmatrix}
z_{rs'} \\ z_{rs}
\end{bmatrix} = R_{1;s'sgc} \begin{bmatrix}
c_r \\ g_r
\end{bmatrix},
\\& \begin{bmatrix}
c_r \\ g_r
\end{bmatrix} = R_{1;gcs's} \begin{bmatrix}
z_{rs'} \\ z_{rs}
\end{bmatrix}.
\end{split}
\end{equation} \label{eq:26}
\\Here, the vectors that are getting transformed are of order 6, i.e., two components corresponding to the two species each of order 3. The two components can be represented by upper and lower indices. Hence, we can also define the mixed Hermite Polynomials:
\begin{flalign}
&\boldsymbol{H}_0 = 1, \label{eq:27}
\\&\boldsymbol{H}_1 = {(\boldsymbol{H}_{1a} \enspace \boldsymbol{H}_{1b})}^T, \label{eq:28}
\\&\textit{\textbf{H}}_{n+1} = \frac{1}{(n+1)!} \enspace S_{(n+1,3)}^{(n+1,2)} (\textit{\textbf{H}}_n \textit{\textbf{H}}_1 - 2n \underline{\underline{I}}^u \textit{\textbf{H}}_{n-1}), \label{eq:29}
\end{flalign}
where $(\underline{\underline{I}}^u)_{ij}^{IJ} = \delta_{IJ}\delta_{ij}; \enspace \enspace I,J = 1,2; \enspace i,j = 1,2,3$ and $S_{(n+1,3)}^{(n+1,2)}$ represents the permutation operator over the indices of both the components.
\\Therefore, using ${(\underline{z}_{rs'}, \enspace \underline{z}_{rs})}^T$ and ${(\underline{c}_r, \enspace \underline{g}_r)}^T$, we can define two sets of Hermite Polynomials. It is observed that the distribution function is the same in rotated and in original axes frame (\ref{app:A7}):
\begin{flalign} \label{eq:eq30}
f_{ss'} = f_{rs}f_{rs'}  = f_{cg}.
\end{flalign}

\section{Discussion}
The description of tensor hermite polynomials and a study of their behavior under scaling, translation and rotation of axes gives us an insight on their applicability in various situations. Since the assumptions in the process were minimal, they provide quite an accurate and generalised computational procedure. This procedure ensures that with the inclusion of each additional tensorial polynomial in the truncated series, the discrepancy between the actual distribution function and the approximated one decreases monotonically. The investigation of these polynomials under scaling gives a condition on temperature which must be satisfied for the collision term of the BTE to be applicable. The polynomials when treated under rotation, provide a simplified way of evaluation.
It should be noted that because of their orthonormality and symmetry, they can be advantageously exploited in simplifying systems other than the one given by Boltzmann Equation.
\\As further research in this area, we plan to work on exploring the properties of the expansion coefficients under various transformation of variables. We also plan to look at the tensorial hermite polynomials in spherical polar coordinates which can be useful for certain systems.

\section*{Acknowledgements}
One of the authors (PM) wishes to thank Tata Consultancy Services, Mumbai for funding this project and Sarthak Bagaria for helping at various stages of the project.

\newpage
\appendix

\section{General Formulae for Hermite Polynomials} \label{app:A1}
\subsection{Grad's Hermite Polynomials, $\boldsymbol{He}_{(n)}$}

General Formulae \cite{1}:
\begin{flalign}
&\nabla_iz_j = \delta_{ij},
\\&{\boldsymbol{\nabla}_j}^n{\boldsymbol{z}_i}^n = {\boldsymbol{\delta}_{ij}}^n,
\\&\boldsymbol{He}_{(n)} = \frac{(-1)^n}{\omega}{\boldsymbol{\nabla}}^n\omega,
\\&\boldsymbol{He}_{(n+1)} = (z_i - {\nabla}_i)\boldsymbol{He}_{(n)}, \label{eq:eqA5}
\\&\boldsymbol{\nabla}_i\boldsymbol{He}_{(n)} = \boldsymbol{\delta}_i\boldsymbol{He}_{(n-1)}, \label{eq:eqA6}
\\&\boldsymbol{He}_{(n)} = {(\boldsymbol{z} - \boldsymbol{\nabla})}^n.\underline{\underline{I}},
\\&\int {\omega \boldsymbol{He}_{(n)i}\boldsymbol{He}_{(n)j}d\boldsymbol{z} = {\boldsymbol{\delta}^n}_{ij}}.
\end{flalign}
where $\boldsymbol{He}_{(n)i}$ is the hermite polynomial with indices $i_1, i_2, i_3, \dots ,i_n$ and $\omega$ represents the normalized weight:
\begin{equation}
\omega(\boldsymbol{z}) = \frac{1}{{(2\pi)}^\frac{N}{2}}e^{-\frac{1}{2}z^2}.
\end{equation}
\\and the following holds \cite{1}
\begin{flalign}
&\nabla_i\omega = -\boldsymbol{z}_i\omega,
\\&\nabla_i(\frac{1}{\omega}) = \frac{\boldsymbol{z}_i}{\omega}.
\end{flalign}
$\boldsymbol{\delta}$ is a second order unit tensor, i.e., the identity matrix and $\boldsymbol{\delta}_i$ represents sum of all terms in which $i$ is attached with $\boldsymbol{\delta}$ as explained in reference \cite{1}.

\subsection{Physicist's Hermite Polynomial, $\boldsymbol{H}_n$}
The relation between physicist's polynomial used in this paper and mentioned in the thesis of Sengupta \cite{8} and the probabilist's polynomial used by Grad \cite{1} is as follows:
\begin{flalign}
&\boldsymbol{H}_n(z) = 2^{\frac{n}{2}}\boldsymbol{He}_{(n)}(\sqrt{2}z),
\\&\boldsymbol{He}_{(n)}(z) = 2^{-\frac{n}{2}}\boldsymbol{H}_n(z/\sqrt{2}).
\end{flalign}
where $\boldsymbol{He}_{(n)}$ is the probabilist's hermite polynomial. Now substituting for $\boldsymbol{He}_{(n)}(z)$ in equation \ref{eq:eqA6} gives:
\begin{equation}
\nabla_i(2^{-\frac{n}{2}}\boldsymbol{H}_n(z/\sqrt{2})) = \boldsymbol{\delta}_i(2^{-\frac{(n-1)}{2}}\boldsymbol{H}_{n-1}(z/\sqrt{2})).
\end{equation}
Replacing $z/\sqrt{2}$ by $z$, we get:
\begin{flalign}
& \nabla_i\boldsymbol{H}_n = 2\boldsymbol{\delta}_i\boldsymbol{H}_{n-1}.
\end{flalign}
Similarly, in equation \ref{eq:eqA5}, substituting for the probabilist's polynomial, we get:
\begin{equation}
\begin{split}
& \quad \quad 2^{-\frac{(n+1)}{2}}\boldsymbol{H}_{n+1} = 2^{-\frac{n}{2}}[\sqrt{2}x_i\boldsymbol{H}_n - \frac{1}{\sqrt{2}}\nabla_i\boldsymbol{H}_n],
\\&or \quad \boldsymbol{H}_{n+1} = (2x_i - \nabla_i)\boldsymbol{H}_n.
\end{split}
\end{equation}

\section{The permutation operator} \label{app:A2}
The permutation operator $S_{(n,m)}$ acts on an argument and replaces the argument by a sum of all possible permutations over n indices for tensors in m-dimensional space \cite{8}.
\\For example, 
\begin{flalign}
&S_{(2,3)}(\boldsymbol{\delta}) = \boldsymbol{\delta}_{ij} + \boldsymbol{\delta}_{ji},
\\&S_{(3,3)}(x\boldsymbol{\delta}) = x_i\boldsymbol{\delta}_{jk} + x_i\boldsymbol{\delta}_{kj} + x_j\boldsymbol{\delta}_{ki} + x_j\boldsymbol{\delta}_{ik} + x_k\boldsymbol{\delta}_{ij} + x_k\boldsymbol{\delta}_{ji}.
\end{flalign}

\section{Proof for the iterative formula (eq. \ref{eq:eq1})} \label{app:A3}
\begin{equation} \label{eq:eqC1}
\begin{split}
&S_{(n+1,3)}[\boldsymbol{H}_{n}\boldsymbol{H}_{1} - 2n\boldsymbol{H}_{n-1}\underline{\underline{I}}]
\\& =\; S_{(n+1,3)}[\boldsymbol{H}_{n}\boldsymbol{H}_{1}] - S_{(n+1,3)}[2n\boldsymbol{H}_{n-1}\underline{\underline{I}}],
\\& =\; [2x_{i_1}S_{(n,3)}(\boldsymbol{H}_n) + 2x_{i_2}S_{(n,3)}(\boldsymbol{H}_n) + 2x_{i_3}S_{(n,3)}(\boldsymbol{H}_n) + \dots] 
\\& \; \; \;- 2n[\delta_{i_1i_2}S_{(n-1,3)}(\boldsymbol{H}_{n-1}) + \delta_{i_1i_3}S_{(n-1,3)}(\boldsymbol{H}_{n-1}) + \dots
\\& \qquad \qquad \delta_{i_2i_1}S_{(n-1,3)}(\boldsymbol{H}_{n-1}) + \delta_{i_2i_3}S_{(n-1,3)}(\boldsymbol{H}_{n-1}) + \dots
\\& \qquad \qquad \dots
\\& \qquad \qquad \dots \qquad + \qquad \dots \qquad + \delta_{i_{n+1}i_n}S_{(n-1,3)}(\boldsymbol{H}_{n-1})].
\end{split}
\end{equation}
Here, $S_{(n,3)}$ the permutation operator explained in the previous section.
\\Since the $H_n$ are the same for any permutations of the indices, we have
\begin{equation}
S_{(n,3)}(\boldsymbol{H}_n) = n!\boldsymbol{H}_n.
\end{equation}
Therefore the above expression is: (putting $H_1 = 2x_i$)
\begin{equation}
\begin{split}
S_{(n+1,3)}[\boldsymbol{H}_{n}\boldsymbol{H}_{1} - 2n\boldsymbol{H}_{n-1}\underline{\underline{I}}] = &n![2x_{i_1}\boldsymbol{H}_n + 2x_{i_2}\boldsymbol{H}_n + \dots]
\\&-2n(n-1)![\delta_{i_1i_2}(\boldsymbol{H}_{n-1}) + \delta_{i_1i_3}(\boldsymbol{H}_{n-1}) + \dots
\\& \qquad \dots 
\\& \qquad \dots \qquad + \qquad \dots \qquad + \delta_{i_{n+1}i_n}(\boldsymbol{H}_{n-1})],
\\= & n![2x_{i_1}\boldsymbol{H}_n + 2x_{i_2}\boldsymbol{H}_n + \dots]
\\&-2n![\boldsymbol{\delta}_{i_1}(\boldsymbol{H}_{n-1}) + \boldsymbol{\delta}_{i_2}(\boldsymbol{H}_{n-1}) + \dots + \boldsymbol{\delta}_{i_{n+1}}(\boldsymbol{H}_{n-1})].
\end{split}
\end{equation}
where the $\boldsymbol{\delta}_i$'s are obtained from the expression in equation \ref{eq:eqA6}. 
\begin{equation}
\begin{split}
S_{(n+1,3)}[\boldsymbol{H}_{n}\boldsymbol{H}_{1} - 2n\boldsymbol{H}_{n-1}\underline{\underline{I}}] = & n![2x_{i_1}\boldsymbol{H}_n + 2x_{i_2}\boldsymbol{H}_n + \dots]
\\&-n![\nabla_{i_1}\boldsymbol{H}_n + \nabla_{i_2}\boldsymbol{H}_n + \dots + \nabla_{i_{n+1}}\boldsymbol{H}_n],
\\= & n![(2x_{i_1}-\nabla_{i_1})\boldsymbol{H}_n + (2x_{i_2}-\nabla_{i_2})\boldsymbol{H}_n + \quad 
\\ &\dots \quad + (2x_{i_{n+1}}-\nabla_{i_{n+1}})\boldsymbol{H}_n].
\end{split}
\end{equation}
Using the relation: $\boldsymbol{H}_{n+1} = (2x_i - {\nabla}_i)\boldsymbol{H}_{n}$ we get:
\begin{equation}
\begin{split}
S_{(n+1,3)}[\boldsymbol{H}_{n}\boldsymbol{H}_{1} - 2n\boldsymbol{H}_{n-1}\underline{\underline{I}}] = & n![\boldsymbol{H}_{n+1} + \boldsymbol{H}_{n+1} + \quad \dots \quad + \boldsymbol{H}_{n+1}],
\\= & (n+1)!\boldsymbol{H}_{n+1}.
\end{split}
\end{equation}

\section{Scalar Inner Product} \label{app:A4}
The scalar inner product of two tensors \textbf{A} and \textbf{B} of \textit{n}-th order is defined as \cite{8}:
\begin{equation}
(\mathbf{A}^{(n)}.\mathbf{B}^{(n)}) = \sum_{i_{1}=1}^{N}\sum_{i_{2}=1}^{N}\dots\sum_{i_{n}=1}^{N}A_{i_1i_2\dots i_n}B_{i_1i_2\dots i_n}
\end{equation}
where N is the number of dimensions in the space under consideration.
\\For example, in 3 dimensions, for two second order tensors (matrices), the scalar inner product will be given as:
\begin{flalign}
&(\mathbf{A}^{(2)}.\mathbf{B}^{(2)}) = A_{11}B_{11} + A_{12}B_{12} + A_{13}B_{13}
\\& \qquad \qquad \qquad + A_{21}B_{21} + A_{22}B_{22} + A_{23}B_{23}
\\& \qquad \qquad \qquad + A_{31}B_{31} + A_{32}B_{32} + A_{33}B_{33}
\end{flalign}


\section{Proof for distribution function in rotated axes frame (eq. \ref{eq:eq30})} \label{app:A6}
\begin{equation}
\begin{split}
f_{ss'} = f_{rs}f_{rs'} &= w(\underline{z}_{rs})w(\underline{z}_{rs'})(\boldsymbol{a}_{nrs},\boldsymbol{H}_{nrs})(\boldsymbol{a}_{mrs'},\boldsymbol{H}_{mrs'}),
\\&= w(\underline{z}_{rs})w(\underline{z}_{rs'})((\boldsymbol{a}_{nrs} \enspace \boldsymbol{a}_{mrs'})^T, (\boldsymbol{H}_{nrs} \enspace \boldsymbol{H}_{mrs'})^T),
\\&= w(\begin{bmatrix} z_{rs'} \\ z_{rs} \end{bmatrix}) (\boldsymbol{\alpha}_{N;rss'}, \textbf{\textit{H}}_{N;rss'}),
\\&= w(\begin{bmatrix} z_{rs'} \\ z_{rs} \end{bmatrix}) (\boldsymbol{\alpha}_{N;rss'}, (\textbf{R}_{N;s'sgc} \enspace \textbf{\textit{H}}_{N;cg})),
\\&= w(\begin{bmatrix} z_{rs'} \\ z_{rs} \end{bmatrix}) ((\boldsymbol{\alpha}_{N;rss'} \enspace \textbf{R}_{N;s'sgc}), \textbf{\textit{H}}_{N;cg}),
\\&= w(\begin{bmatrix} z_{rs'} \\ z_{rs} \end{bmatrix}) ((\textbf{R}_{N;gcs's} \enspace \boldsymbol{\alpha}_{N;rs's}), \textbf{\textit{H}}_{N;cg}),
\\&= w(\begin{bmatrix} c_{r} \\ g_{r} \end{bmatrix}) (\boldsymbol{\beta}_{N;gc}, \textbf{\textit{H}}_{N;cg}) \enspace = \enspace f_{cg}.
\end{split}
\end{equation}
where, $\alpha$ is a tensor of rank $N=m+n$ and order 6 and $\beta$ is the corresponding set of rotated mixed tensor expansion coefficients. $\textbf{R}_{N;s'sgc}$ is the N-th rank rotation tensor.

\section{Proof for hermite polynomials under translation (eq. \ref{eq:eq19} and \ref{eq:eq20})} \label{app:A7}
This proof can be done by principle of mathematical induction. Substituting for the first few values of $n$ or $p$, we observe that the expressions hold true. Assuming that the expression
\begin{equation}
\boldsymbol{H}_n^0 = \frac{1}{n!}S_{(n,3)}[\sum_{p=0}^n (_p^n) \boldsymbol{H}_p^r [2(\underline{z}_a-\underline{z}_{00})]^{t(n-p)}],
\end{equation}
holds true for $n$ then proving that this implies it also holds true for $n+1$ shows that this is true in general.
From equation (1), we have
\begin{equation}
\boldsymbol{H}_{n+1}^0 = \textbf{H}_n^0\textbf{H}_1^0 - 2n\textbf{H}_{n-1}^0\underline{\underline{I}},
\end{equation} since the $\textbf{H}_n's$ are symmetric in indices, $S_{(n+1,3)}$ and $(n+1)!$ cancel.
\\Subsituting for the series expression of $\boldsymbol{H}_n^0$

{ \allowdisplaybreaks
\begin{align*}
\boldsymbol{H}_{n+1}^0 & = \sum_{p=0}^n (_p^n)\boldsymbol{H}_p^r{[2(\underline{z}_a -\underline{z}_{00})]}^{t(n-p)}2(\underline{z}-\underline{z}_{00}) - 2n\sum_{p=0}^{n-1} (_p^{n-1})\boldsymbol{H}_p^r{[2(\underline{z}_a -\underline{z}_{00})]}^{t(n-p-1)}\underline{\underline{I}},
\\& = \sum_{p=0}^n (_p^n)\boldsymbol{H}_p^r{[2(\underline{z}_a -\underline{z}_{00})]}^{t(n-p)}2(\underline{z}-\underline{z}_{00})
\\& \qquad \qquad - 2\sum_{p=0}^{n-1} (_{p+1}^{n})(p+1)\boldsymbol{H}_p^r{[2(\underline{z}_a -\underline{z}_{00})]}^{t(n-p-1)}\underline{\underline{I}},
\\& = {[2(\underline{z}_a -\underline{z}_{00})]}^{t(n)}2(\underline{z}-\underline{z}_{00}) + \sum_{p=1}^n (_p^n)\boldsymbol{H}_p^r{[2(\underline{z}_a -\underline{z}_{00})]}^{t(n-p)}2(\underline{z}-\underline{z}_{00})
\\& \qquad \qquad - 2\sum_{p=1}^{n} (_{p}^{n})(p)\boldsymbol{H}^r_{p-1}{[2(\underline{z}_a -\underline{z}_{00})]}^{t(n-p)}\underline{\underline{I}},
\\& = {[2(\underline{z}_a -\underline{z}_{00})]}^{t(n+1)} + {[2(\underline{z}_a -\underline{z}_{00})]}^{t(n)} \boldsymbol{H}^r_{1} + \sum_{p=1}^n (_p^n){[2(\underline{z}_a -\underline{z}_{00})]}^{t(n-p)}
\\& \qquad \qquad \quad [\boldsymbol{H}_p^r\boldsymbol{H}^r_{1} + \boldsymbol{H}_p^r2(\underline{z}_a-\underline{z}_{00}) - 2p\boldsymbol{H}^r_{p-1} \underline{\underline{I}}],
\\& = {[2(\underline{z}_a -\underline{z}_{00})]}^{t(n+1)} + {[2(\underline{z}_a -\underline{z}_{00})]}^{t(n)} \boldsymbol{H}^r_{1} + \sum_{p=1}^n (_p^n)\boldsymbol{H}^r_{p+1}{[2(\underline{z}_a -\underline{z}_{00})]}^{t(n-p)}
\\& \qquad \qquad + \sum_{p=1}^n (_p^n)\boldsymbol{H}_p^r{[2(\underline{z}_a -\underline{z}_{00})]}^{t(n-p+1)},
\\& = {[2(\underline{z}_a -\underline{z}_{00})]}^{t(n+1)} + {[2(\underline{z}_a -\underline{z}_{00})]}^{t(n)} \boldsymbol{H}^r_{1} + \sum_{p=1}^n (_p^n)\boldsymbol{H}^r_{p+1}{[2(\underline{z}_a -\underline{z}_{00})]}^{t(n-p)}
\\& \qquad \qquad + \sum_{p=0}^{n-1} (_{p+1}^n)\boldsymbol{H}^r_{p+1}{[2(\underline{z}_a -\underline{z}_{00})]}^{t(n-p)},
\\& = {[2(\underline{z}_a -\underline{z}_{00})]}^{t(n+1)} + (n+1){[2(\underline{z}_a -\underline{z}_{00})]}^{t(n)} \boldsymbol{H}^r_{1}
\\& \qquad \qquad + \sum_{p=1}^{n-1} (_{p+1}^{n+1})\boldsymbol{H}^r_{p+1}{[2(\underline{z}_a -\underline{z}_{00})]}^{t(n+1-(p+1))} + \boldsymbol{H}^r_{n+1},
\\& = {[2(\underline{z}_a -\underline{z}_{00})]}^{t(n+1)} + (n+1){[2(\underline{z}_a -\underline{z}_{00})]}^{t(n)} \boldsymbol{H}^r_{1}
\\& \qquad \qquad + \sum_{p=0}^{n-2} (_{p}^{n+1})\boldsymbol{H}_p^r{[2(\underline{z}_a -\underline{z}_{00})]}^{t(n+1-p)} + \boldsymbol{H}^r_{n+1},
\\& = \sum_{p=0}^{n+1} (_{p}^{n+1})\boldsymbol{H}_p^r{[2(\underline{z}_a -\underline{z}_{00})]}^{t(n+1-p)}.
\end{align*}
}

\newpage


\end{document}